\documentclass[twocolumn,nodate,showpacs,preprintnumbers,amsmath,amssymb,aps,prd]{revtex4}
\usepackage{graphicx}
\usepackage{bm}  
\def\lqc#1{Loop Quantum Cosmology#1 (LQC#1)\gdef\lqc{LQC}} 
\def\wdw#1{Wheeler-DeWitt#1 (WDW#1)\gdef\wdw{WDW}}       

\begin{document}

\title{Numerical Analysis of the Big Bounce in Loop Quantum Cosmology}

\author{Pablo Laguna}
\altaffiliation[Also at ]{IGPG, CGWP, Physics Department.}
\affiliation{Department of Astronomy and Astrophysics\\
Penn State University\\
University Park, PA 16802}             

\begin{abstract}
\noindent 
Loop quantum cosmology homogeneous models with a massless scalar field show that the big-bang singularity can
be replaced by a big quantum bounce. To gain further insight on the nature of this bounce, 
we study the semi-discrete loop quantum gravity Hamiltonian constraint equation from
the point of view of numerical analysis. 
For illustration purposes, 
we establish a numerical analogy between the quantum bounces and reflections in finite difference discretizations
of wave equations triggered by the use of nonuniform grids or, equivalently, reflections found when
solving numerically wave equations with varying coefficients.
We show that the bounce is closely related to the
method for the temporal update of the system and demonstrate
that explicit time-updates in general yield bounces. 
Finally, we present an example of an implicit time-update devoid of bounces 
and show back-in-time, deterministic evolutions 
that reach and partially jump over the big-bang singularity.  
\end{abstract}  


\maketitle

\lqc{} provides a framework to explore the dynamics of the Universe deep inside the Plank 
regime~\cite{Ashtekar:2004eh,Rovelli:2004tv,Thiemann:2002nj}.
In \lqc{,} quantum evolution is no longer dictated by the second order \wdw{} differential equation,
but instead the dynamics is governed by a second order finite difference equation, the so-called
\lqc{} Hamiltonian constraint. The \wdw{} and \lqc{} equations differ only when quantum geometry effects
cannot be ignored. A focus of \lqc{} has been the problem of the resolution of the big-bag singularity.
Early studies~\cite{2001PhRvL..86.5227B,2000CQGra..17.1489B,2000CQGra..17.1509B} 
showed that the \lqc{} difference 
equation remains deterministic and well defined at the classical singularity,
thus allowing evolution that {\it jumps} over the big-bang. 
Recent studies~\cite{2006PhRvL..96n1301A,2006PhRvD..73l4038A,Ashtekar:2006wn} of spatially flat, isotropic
models with a massless scalar field $\phi$ have also shown that, 
under the \lqc{} framework, the big-bang singularity is replaced by a big quantum bounce.

For these models, 
the \lqc{} Hamiltonian constraint can be written as~\cite{Ashtekar:2006wn}
\begin{eqnarray}
\label{eq:lqc1}
\nonumber
\partial^2_\phi\Psi(v) &=& \frac{3}{2}\kappa\,B^{-1}(v)\lbrace\\ \nonumber
&&C(v+2v_o)\left[\Psi(v+4v_o)-\Psi(v)\right]-\\ 
&&C(v-2v_o)\left[\Psi(v)-\Psi(v-4v_o)\right]\rbrace\,,
\end{eqnarray}
where
$B(v) = 3^3|v|\left||v+v_o|^{1/3}-|v-v_o|^{1/3}\right|^3$ 
is related to the eigenvalues of the inverse volume operator~\cite{Ashtekar:2003hd} and 
$C(v) = |v|\left||v+v_o|-|v-v_o|\right|/4$. Units are such that $\kappa \equiv 8\pi G$.
In Eq.~(\ref{eq:lqc1}), it is implied that $\Psi$ also depends on $\phi$. 
The discrete independent variable 
$v$ is the eigenvalue of the volume operator, with the
discretization scale $v_o$ fixed by the {\it area gap}.
The scalar field $\phi$ is used as an internal time.
Starting with a state that is semi-classical at late times,
evolutions backwards in time using
Eq.~(\ref{eq:lqc1}) yield a turning point, a quantum bounce, before 
reaching the classical singularity  at $v=0$~\cite{Ashtekar:2006wn}. The turning point takes place                    
when the total energy density reaches a critical value $\rho_{crit} < \rho_{Planck}$.

The goal of this {\it Letter} is to gain further insight, from a numerical analysis point
of view, about the nature of this quantum bounce. We will show that the observed turning
point is strongly tied to the method to carry out the temporal updating of the system.   
Specifically, explicit time-update of the semi-discrete Eq.~(\ref{eq:lqc1}),
such as the 4th-order Runge-Kutta updating used in 
Refs.~\cite{2006PhRvL..96n1301A,2006PhRvD..73l4038A,Ashtekar:2006wn},
will yield a bounce. One can draw a numerical analogy in which the bounce can be understood in the context of spurious 
reflections from finite difference discretizations of wave equations in
nonuniform grids~\cite{frank-spurious} or equivalently from numerical solutions to wave equations with varying 
coefficients.
Bounces due to discreteness have been also obtained 
in the study of Hawking radiation on a falling lattice\cite{2000PhRvD..61b4017J}.

Motivated by understanding the differences between various ambiguities in the constraints for the
full theory, we will also show that the bounce could be avoided by introducing 
an implicit time-update, which is formally equivalent to adding ad hoc higher order 
terms to the \lqc{} Hamiltonian constraint.
With this implicit update, it is possible to evolve backwards in time 
a state that reaches the classical big-bang singularity. At that point, the wavefunction $\Psi$ gets 
partially reflected and partially transmitted through the singularity.
The resulting evolution is deterministic and thus also resolves the problem of the big-bang singularity.

We first introduce the standard mesh-index notation used in finite differences, namely
$v_j = j\,\Delta v$ with $\Delta v=4\,v_o$.
In this notation, the semi-discrete Eq.~(\ref{eq:lqc1}) takes the form
\begin{equation}
\label{eq:lqc2}
\partial^2_\phi\Psi_j = \frac{3}{2}\kappa\widetilde{B}^{-1} D_v(\widetilde{C}D_v \Psi_j)
\end{equation}
where $\widetilde{B} = (3|v|^{2/3}\widetilde{D}_v |v|^{1/3})^3/|v|$ and
$\widetilde{C} = |v|\widetilde{D}_v|v|$ with
$D_v f_j \equiv (f_{j+1/2}-f_{j-1/2})/\Delta v$ and
$\widetilde{D}_v f_j \equiv (f_{j+1/4}-f_{j-1/4})/(\Delta v/2)$ discrete finite difference operators.
In the limit $\Delta v\rightarrow 0$, both of these 
operators become $\partial_v$ as well as $\widetilde{B}\rightarrow 1/|v|$ 
and $\widetilde{C} \rightarrow |v|$.
Also in that limit, Eq.~(\ref{eq:lqc2}) becomes 
$\partial^2_\phi\Psi = 3\kappa v/2 \partial_v (v\partial_v \Psi)$; that is, one recovers the \wdw{} equation
in the $v\propto a^3$ representation where $a$ is the scale factor. 

To simplify our analysis, we will work with a version of Eq.~(\ref{eq:lqc2}) in which 
$\widetilde{B} \approx 1/|v|$ and $\widetilde{C} \approx |v|$; that is,
$\partial^2_\phi\Psi_j =  (3\kappa |v|/2) D_v(|v|D_v \Psi_j)$. 
The differences between this equation and Eq.~(\ref{eq:lqc2}) are of $O(\Delta v^2)$ and
do not affect our conclusions regarding the big quantum bounce.
In addition, since ultimately the mean trajectories of states in the $\phi-v$ plane are 
given by the characteristic of $\partial^2_\phi\Psi_j =  (3\kappa |v|/2 D_v)(|v|D_v \Psi_j)$, 
we will concentrate our attention on its
principal part, namely $\partial^2_\phi\Psi_j =  c^2 D^2_v\Psi_j$  where
$c=\sqrt{3\kappa/2}|v|$. 

The starting point of our analysis is applying a time-Fourier
transformation $\Psi_j(\phi) = \int^\infty_{-\infty} e^{-i\omega\phi}\widehat{\Psi}_j(\omega)d\omega$.
Thus, $-\omega^2\widehat{\Psi}_j =  c^2 D^2_v\widehat{\Psi}_j$. Next, we insert the plane wave solutions
$\widehat{\Psi}_{j\pm 1} = e^{\pm ik\Delta v}\widehat{\Psi}_j$ with wavenumber $k$. 
For $\Delta v\, \omega/2\,c\le 1$, this yields
the following dispersion relationship
\begin{equation}
\label{eq:dispersion}
 \frac{\Delta v\, \omega}{2\,c}= \pm\sin{\left(\frac{\Delta v\,k}{2}\right)}\,.
\end{equation}
For $\Delta v\, \omega/2\,c > 1$, the wave is exponentially damped.
From this dispersion relation, the group velocity $V_g=d\omega/dk$ 
is given by 
\begin{equation}
\label{eq:vg}
V_g = \pm c\, \cos{\left(\frac{\Delta v\, k}{2}\right)} 
= \pm c\left[1-\left(\frac{\Delta v\, \omega}{2\,c}\right)^2\right]^{1/2}\,.
\end{equation}
It is then clear that the states, solutions of the semi-discrete equation,
will have different dynamics from those of the continuum \wdw{} equation.
\wdw{} states have a group velocity $\pm c$ and follow characteristics 
$v = v_*e^{\pm\sqrt{3\kappa/2}(\phi-\phi_*)}$ with $v_*$ and $\phi_*$ integration constants.
On the other hand, \lqc{} states have group velocity (\ref{eq:vg}) $|V_g| \le c$
and mean trajectories $v = \xi/2[ e^{\pm\sqrt{3\kappa/2}(\phi-\phi_*)}+(v_b/\xi)^2e^{\mp\sqrt{3\kappa/2}(\phi-\phi_*)}]$,
with $\xi = v_*+\sqrt{v_*^2-v_b^2}$ and $v_b = \Delta v \,\omega/\sqrt{6\kappa}$.
The value $v_b$ is the location where 
the group velocity vanishes. This is where the characteristics reverse direction,
the big quantum bounce. The condition $(\Delta v\, \omega / 2\,c)^2 = 1$  is precisely the condition 
$\rho/\rho_{crit} = 1$ for the quantum bounce derived in Ref.~\cite{Ashtekar:2006wn} after identifying 
$\omega$ with $P_\phi$, the momentum conjugate to ``time." 

Applying an identical analysis to the version of \lqc{} Hamiltonian 
constraint in Ref.~\cite{2006PhRvD..73l4038A,2006PhRvL..96n1301A}, one arrives to 
Eq.~(\ref{eq:vg}) with $v$ replaced by 
the triad $p$ and setting $c=\sqrt{2\kappa/3}|p|$. The location of the bounce is now at
$p_b = \Delta p\, \omega / \sqrt{8\kappa/3}$. Since $v \propto p^{3/2}\propto a^3$, when
translated to the $a$ representation, the 
turning points $v_b$ and $p_b$ occur at different values of $a$, or equivalently, for 
different total energy densities $\rho_{crit}$.
This ``coordinate transformation" was the key ingredient in Ref.~\cite{Ashtekar:2006wn}
for having the bounce occurring at $\rho_{crit} < \rho_{Planck}$.
 
It is very important to emphasize that the derivation of $V_g$ and its turning point condition
did not depend on any specific time-update 
(e.g. Staggered Leap-Frog, Crank-Nicholson, Runge-Kutta, etc). 
The only two assumptions used were
the ``spatial" discretization in Eq.~(\ref{eq:lqc1}) or (\ref{eq:lqc2}), 
i.e. centered second order accurate finite differences, 
and that the time-update is {\it explicit}; that is, the 
numerical approximation to $\partial^2_\phi\Psi_j$ was assumed to depend only on $j$. 
In other words, the l.h.s.
of Eq.~(\ref{eq:lqc1}) or (\ref{eq:lqc2}) only depends on $v_j$. 
From the \lqc{} point of view, the explicit time-update arises because
the inverse scale factor operator used in deriving Eq.~(\ref{eq:lqc1}) is diagonal.
Also important is to keep in mind that, since the truncation errors 
from approximating 
$\widetilde{B}\approx 1/|v|$ and $\widetilde{C} \approx |v|$ are small and only become relevant
in the immediate vicinity of the classical singularity, 
the bounce observed from the \lqc{} equation would be identical to that from solving numerically the \wdw{} equation
with second order center difference approximations
and explicit time-integration. 
In this regard, what distinguishes \lqc{} and \wdw{} equations is the 
specific forms of the spatial operators, continuum operators for \wdw{} and discrete for \lqc{}.

A possible way of avoiding the bounce in the discrete \wdw{} equation is to
perform the coordinate transformation $\alpha \propto \ln{a} \propto \ln{v^{1/3}}$. 
With this transformation, 
the \wdw{} equation becomes $\partial^2_\phi\Psi = c^2 \partial^2_\alpha\Psi$
with $c=\sqrt{\kappa/6}$. Using a uniform mesh in $\alpha$, 
the group velocity in this case is identical to (\ref{eq:vg})
with $\Delta v$ replaced by $\Delta \alpha$.
Thus, provided the initial state satisfies
$\Delta \alpha\, \omega/2\,c < 1$, the group velocity will not vanish. 
The same coordinate transformation can be applied to the \lqc{} Eq.~(\ref{eq:lqc1}); however,
because the \lqc{} theory requires keeping $\Delta v$ constant, 
the mesh in the $\alpha$ coordinate will not be uniform. 
It is in this context that one can
view the bounce as spurious reflections due to nonuniform grids~\cite{frank-spurious}.

So far, we have only revisited the semi-discrete \lqc{} Eq.~(\ref{eq:lqc1}) under the
numerical analysis eyepiece. In particular, we have demonstrated that the nature of the
quantum bounce is not entirely due to the specific form of the ``spatial" operator in
the r.h.s. of the equation, which is the main difference between the
\lqc{} and \wdw{} equations. Also very important, and closely intertwined for the appearance of 
the bounce, is the explicit time-update implied by l.h.s. of the difference equation.
If time-updates play a crucial role, one could 
ask whether other time-updates could radically change the dynamics,
and in particular the bounce, 
while keeping fixed the spatial discretization. The answer is affirmative. 

Let us consider the following modification to Eq.~(\ref{eq:lqc2}),
\begin{equation}
\label{eq:lqc3}
M^2_v\partial^2_\phi\Psi_j = \frac{3}{2}\kappa\widetilde{B}^{-1} D_v(\widetilde{C}D_v \Psi_j)\,
\end{equation}
where $M_v f_j = (f_{j+1/2}+f_{j-1/2})/2$. 
This is an example of an {\it implicit} time update 
belonging to the
class of Keller-Preissman {\it box} schemes~\cite{frank-spurious}. 
When expanded, the l.h.s. of the equations reads
$(\partial^2_\phi\Psi_{j+1}+2\,\partial^2_\phi\Psi_{j}
+\partial^2_\phi\Psi_{j-1})/4$.
Which means that one cannot update the value of $\Psi_j$ without the
updated values of $\Psi_{j\pm 1}$.
Implicit time-updates could arise if one considers non-diagonal 
terms of the inverse volume operator~\cite{2001PhRvL..86.5227B,2001PhRvD..64h4018B,Bojowald:2005ui}. 
Although these non-diagonal terms are likely to only become relevant near the classical singularity.

An interesting observation is that Eq.~(\ref{eq:lqc3}) can be rewritten as
\begin{equation}
\label{eq:lqc4}
\partial^2_\phi\Psi_j =
\frac{3}{2}\kappa\widetilde{B}^{-1} D_v(\widetilde{C}D_v \Psi_j)
+ \left(\frac{\Delta v}{2}\right)^2D^2_v \partial^2_\phi\Psi_{j}\,.  
\end{equation}
In the semi-classical regime, where $O(\Delta v^2)$ terms can be ignored, 
Eq.~(\ref{eq:lqc4}) exhibits the same explicit temporal update as the standard \lqc{} equation.
On the other hand, near the classical singularity, one could apply $D^2_v$ to Eq.~(\ref{eq:lqc4}) and 
use the resulting equation to eliminated, recursively, in the r.h.s of Eq.~(\ref{eq:lqc4}) the terms
with $\partial^2_\phi \Psi_{j}$ in favor of higher order, spatial finite difference terms.
However, in order to arrive to an equation that completely avoids the bounce, one would have to
include enough higher order terms, that are effectively equivalent to an implicit update. 

To derive the dispersion relation and group velocity for the modified
\lqc{} difference equation,
we concentrate once again on the principal of Eq.~(\ref{eq:lqc3}) 
and perform a time-Fourier transform. One obtains 
$-\omega^2\,M^2_v\widehat{\Psi}_j =  c^2 D^2_v\widehat{\Psi}_j$
with $c=\sqrt{3\kappa/2}|v|$. 
After substitution of the plane-wave solution,
the dispersion relation in this case takes the form
\begin{equation}
\label{eq:dispersion2}
 \frac{\Delta v\, \omega}{2\,c}= \pm\tan{\left(\frac{\Delta v\,k}{2}\right)}\,.
\end{equation}
From which the group velocity reads,
\begin{equation}
\label{eq:vg2}
V_g = \pm c\, \sec^2{\left(\frac{\Delta v\, k}{2}\right)} 
= \pm c\left[1+\left(\frac{\Delta v\, \omega}{2\,c}\right)^2\right]\,.
\end{equation}
Clearly this type of time-update does not have a bounce or turning point. 
The group velocity does not vanish for $|v|\ne 0$.
The characteristics in this case are $v=[(v_*^2+v_b^2)e^{\sqrt{6\kappa}(\phi-\phi_*)}-v_b^2]^{1/2}$.
At $v=0$, the group velocity becomes infinite because $c=0$.

\begin{figure}
\scalebox{0.4}{\includegraphics{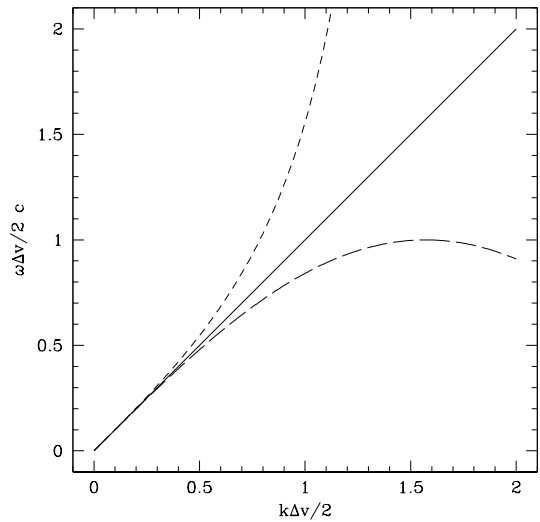}}
\caption{Dispersion relation at an
arbitrary point $|v|\ne 0$ for the \wdw{} equation (solid line),
the \lqc{} equation with explicit time-update (long dashed line) and
the modified \lqc{} equation with implicit time-update (short dashed line).} 
\label{fig:dispersion}
\end{figure}

Fig.~\ref{fig:dispersion} shows the dispersion relation at an
arbitrary point $|v|\ne 0$ for the \wdw{} equation (solid line),
the \lqc{} equation with explicit time-update (long dashed line) and
the modified \lqc{} equation with implicit time-update (short dashed line).  
It is evident that for explicit time-updates, there always exists
a frequency for which the group velocity 
$V_g = d\omega/dk$ vanishes 
and triggers a mode with the opposite group velocity. 
The reason for the total reflection can also be understood in terms of 
the {\it left}- and {\it right}-moving fundamental modes. 
These modes couple because Eq.~(\ref{eq:lqc3}) is a difference 
equation with non-constant coefficients~\cite{frank-spurious}.
The coupling also occurs, even in the case of
constant coefficients, if the mesh is non-uniform or discontinuous~\cite{frank-spurious}.
The coupling is such that a pure mode will be completely reflected 
at the point where $V_g$ vanishes~\cite{Laguna}. 
Also evident in Fig.~\ref{fig:dispersion} is that implicit time-updates
of the type in Eq.~(\ref{eq:lqc3}) do not trigger reflections
because their dispersion relation is monotonic.

\begin{figure}
\scalebox{0.4}{\includegraphics{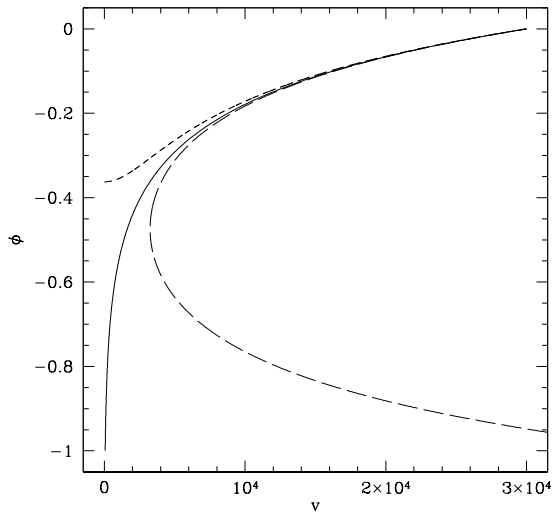}}
\caption{An example of a characteristics of the continuum 
\wdw{} equation (solid line),
the \lqc{} equation with explicit time-update (long dashed line) and
the \lqc{} equation with implicit time-update (short dashed line).Constants of integration
were chosen so the three characteristics have the same starting point when evolved
backwards in time}
\label{fig:charact}
\end{figure}

Fig.~\ref{fig:charact} depicts an example of the characteristics of the continuum 
\wdw{} equation (solid line), the \lqc{} equation with explicit time-update (long dashed line) and
the modified \lqc{} equation with implicit time-update (short dashed line). The constants of integration
were chosen so the three characteristics have the same starting point when evolved
backwards in time.

\begin{figure}
\scalebox{0.35}{\rotatebox{-90}{\includegraphics{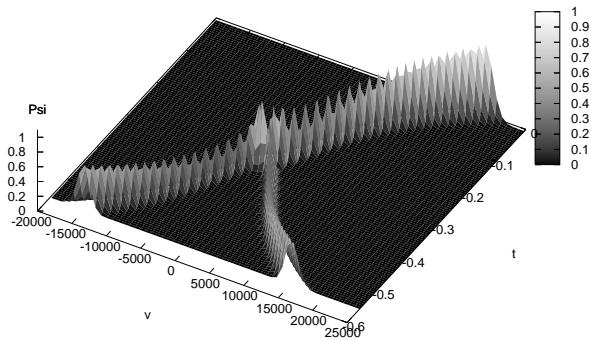}}}
\caption{Evolution backwards in time of a coherent state using
an implicit time update.
The initial data at $\phi^*= 0$ is centered at $v^*=2\times10^4\,v_o$ and 
has momentum $P_\phi = 10^4$.}
\label{fig:shallow}
\end{figure}   

Fig.~\ref{fig:shallow} shows the evolution of a coherent state solution to
Eq.~(\ref{eq:lqc3})
using a mid-point trapezoidal time-update.
The initial data at $\phi^*= 0$ is centered at $v^*=2\times10^4\,v_o$, has
momentum $P_\phi = 10^4$ and is given by 
$\Psi = \exp{[-(Z/\sigma)^2 + i\,P_\phi\,Z]}$
where 
$Z \equiv \phi - \phi^* \pm \sqrt{2/3\,\kappa}\ln{(v/v^*)}$
and 
$\sigma =  \sqrt{2/3\,\kappa}\Delta P_\phi/P_\phi$ with 
$\Delta P_\phi/P_\phi=7.5\times 10^{-2}$. With these parameters, if an explicit time-update
were to be used, the bounce would take place
at $v_b = 3257\,v_o$.

At early times when $v \gg v_b$, the mean trajectory of the state follows the semi-classical trajectory
$v=v_*e^{\sqrt{3\kappa/2}(\phi-\phi_*)}$ until $v$ is comparable to $v_b$. At this point, instead of
approaching a bounce, the state evolves towards the singularity, reaching it in a finite time. 
As observed from Fig.~\ref{fig:shallow}, when the wavefunction reaches $v=0$,
it gets partially reflected and partially transmitted across the classical singularity. 
This partial reflection and transmission is due to the coupling between {\it left}- and {\it right}-moving
modes at $v=0$~\cite{Laguna}. 

Applying tools commonly used in numerical analysis, we have investigated those aspect in the
\lqc{} finite difference equation related to the
bounce observed in homogeneous models of massless scalar fields. 
We find that, because the semi-discrete \lqc{} equation has non-constant coefficients,
explicit time-updates will typically couple the {\it left}- and {\it right}-moving fundamental modes 
and trigger reflections at the point where the group velocity vanishes. 
Explicit time integrations in \lqc{} are tied 
to having a diagonal operator in the gravitational part 
of the matter Hamiltonian for models in which 
one uses a single matter field as internal time~\cite{2002PhRvD..66j4003B}.
In addition, we have investigated ad hoc
modifications to the standard \lqc{} finite difference equation that avoid the big quantum bounce
while preserving a deterministic evolution across the big-bang singularity.
The changes focused on replacing the
explicit time update with an implicit scheme. 
Our numerical experiments showed that a semi-classical state at late times can be evolved backwards in time and
reach the classical singularity in a finite time. At that point, the state gets partially transmitted.  
An interesting implication of this result is the outcome from an evolution forward in time
of a {\it right}-moving (increasing $v$ direction) 
state initially at the ``other side" of the classical singularity. 
When the state reaches the classical singularity, 
it will yield a reflected wavefunction that remains on the other side of the singularity and a
transmitted wavefunction emerging across the classical big bang into the physical sector.

Our analysis and numerical experiments were aimed at investigating the features in 
\lqc{} homogeneous models with a massless scalar field that are related to the occurrence of a bounce. 
A complementary investigation, based on effective perturbation theory 
around a free scalar model, can be found in~\cite{Bojowald:2006gr}. From a 
general perspective, our findings agree for dynamics which allows wave 
packets to reach small scales.
Additional studies are needed to investigate whether more complicated models involving 
the {\it issue of time} could introduce modification similar to those considered here,
in particular instances in which the \lqc{} difference equation
includes non-diagonal elements of the inverse volume operator.
Those terms could have the potential of playing an important role near the big-bang singularity,
significantly modifying or even preventing the big quantum bounce.

\begin{acknowledgments} 
Thanks to A. Ashtekar, M. Bojowald, J. Hartle, D. Marolf, T. Pawlowski and P. Singh for
helpful conversations, their comments and suggestions.
This work was supported by NSF grants PHY-0244788 and PHY-0555436.
CGWP is supported
by the NSF under cooperative agreement PHY-0114375. 
Thanks also to the hospitality of KITP, supported by
NSF PHY-9907949, where part of this work was completed.
\end{acknowledgments}

\bibliographystyle{apsrev}

\end{document}